\documentclass[a4paper]{article}
\usepackage{spconf,amsmath,graphicx,amssymb,delarray,subfigure,verbatim, bbding}
\usepackage{multirow}
\usepackage{INTERSPEECH2020}

\title{The IOA System for Deep Noise Suppression Challenge \\ using a Framework Combining Dynamic Attention and Recursive
Learning}

\name{Andong Li$^{1,2}$, Chengshi Zheng$^{1,2}$, Renhua Peng$^{1,2}$, Linjuan Cheng$^{1, 2}$, and Xiaodong Li$^{1,2}$}

\address{
	$^1$ Key Laboratory of Noise and Vibration Research, Institute of Acoustics, Chinese Academy of Sciences, Beijing, China\\
	$^2$University of Chinese Academy of Sciences, Beijing, China}
\email{\{liandong, cszheng, pengrenhua, chenglinjuan, lxd\}@mail.ioa.ac.cn}

\begin{document}
\hyphenpenalty=7000
\tolerance=1000
\maketitle
\begin{abstract}
This technical report describes our system that is submitted to the Deep Noise Suppression Challenge and presents the results for the non-real-time track. To refine the estimation results stage by stage, we utilize recursive learning, a type of training protocol which aggravates the information through multiple stages with a memory mechanism. The attention generator network is designed to dynamically control the feature distribution of the noise reduction network. To improve the phase recovery accuracy, we take the complex spectral mapping procedure by decoding both real and imaginary spectra. For the final blind test set, the average MOS improvements of the submitted system in noreverb, reverb, and realrec categories are 0.49, 0.24, and 0.36, respectively.

\end{abstract}

\section{Background}
In the INTESPEECH 2020 Deep Noise Suppression (DNS) Challenge, it aims to promote further research towards monaural speech enhancement. When given extensive and representative datasets to train the network, the model is expected to effectively extract the target speech information from the noisy signals and also obtain relatively satisfactory performance in both objective and subjective evaluation. Two tracks are provided, namely real-time (RT) track and non-real-time (NRT) track. For the former, the computational complexity is partly restricted. For the latter, the complexity requirement is relaxed, which makes it possible to explore the network with more complicated topology. More information about the challenge can be referred to~{\cite{DNSChallenge2020}}. In our submitted system, the innovations are threefold: First, recursive learning is utilized, where the model is progressively trained via unfolding the network for multiple stages and the correlation between stages is bridged through a memory mechanism; Second, the dynamic attention mechanism is designed, where a separate unsupervised attention generator network is utilized to adjust the feature distribution of the noise suppression network; Third, complex spectral mapping scheme is utilized to boost the phase recovery.

\section{System description}
\label{system-description}
\begin{figure}[t]
	\centering
	\centerline{\includegraphics[width=\columnwidth]{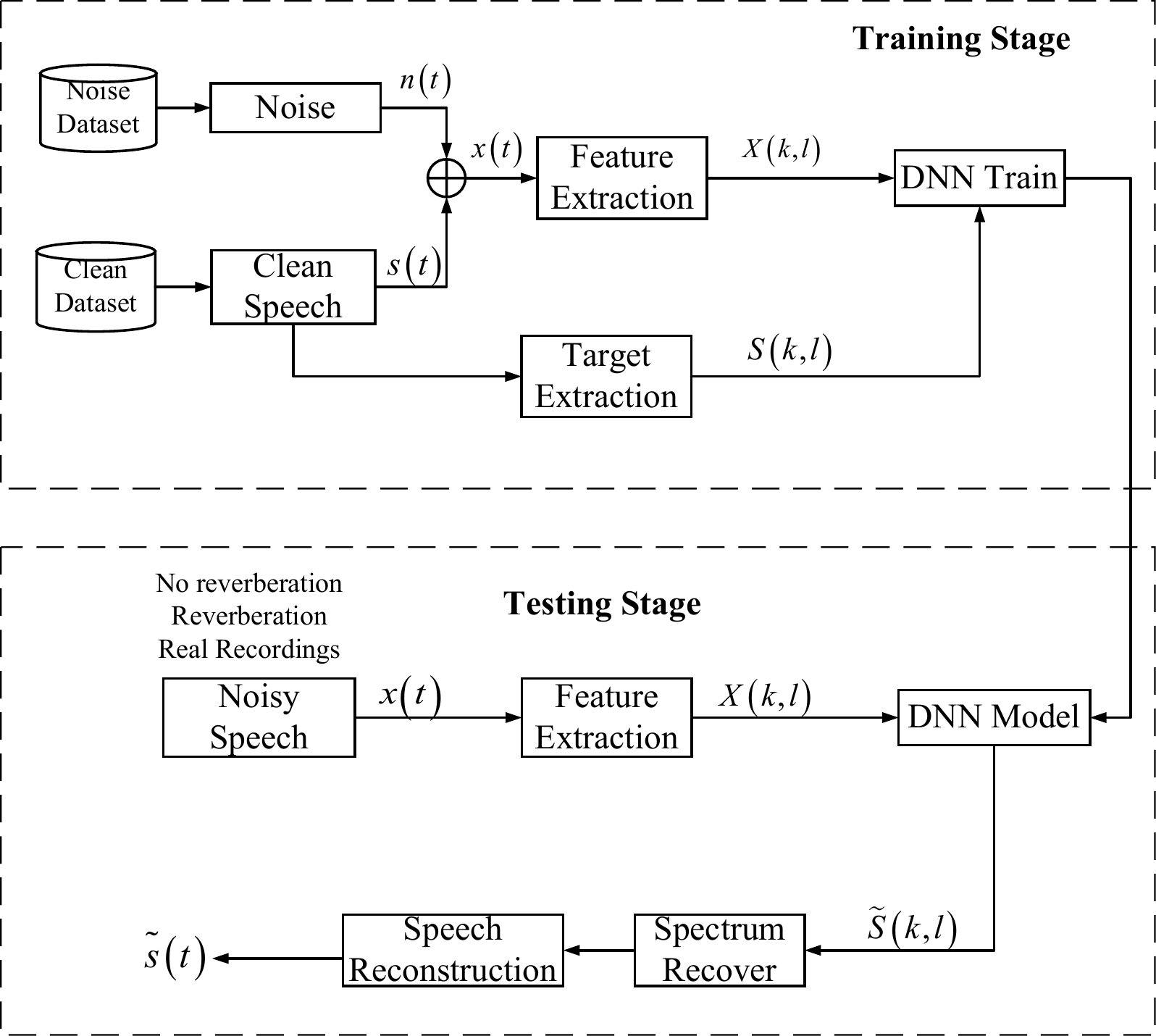}}
	\caption{The flowchart of the system.}
	\label{fig:flow-chart}
\end{figure}
This section describes the system flowchart and architecture in detail. The system flowchart is given in Fig.~{\ref{fig:flow-chart}}. During the training stage, a multitude of training pairs, including extracted noisy features and their corresponding targets, are sent to the network for training until the process has been converged, In the test stage, the testing samples are sent to the network to obtain the enhanced spectrum. Finally, the time-domain signals are reconstructed with the overlap-add technique. Note that three types of testing samples are provided in the DNS Challenge, including synthesized noisy samples without reverberation (noreverb), synthesized noisy samples with reverberation (reverb), and real-recordings (realrec).

\begin{figure*}[t]
	\setlength{\abovecaptionskip}{0.235cm}
	\setlength{\belowcaptionskip}{-0.1cm}
	\centering
	\centerline{\includegraphics[width=155mm]{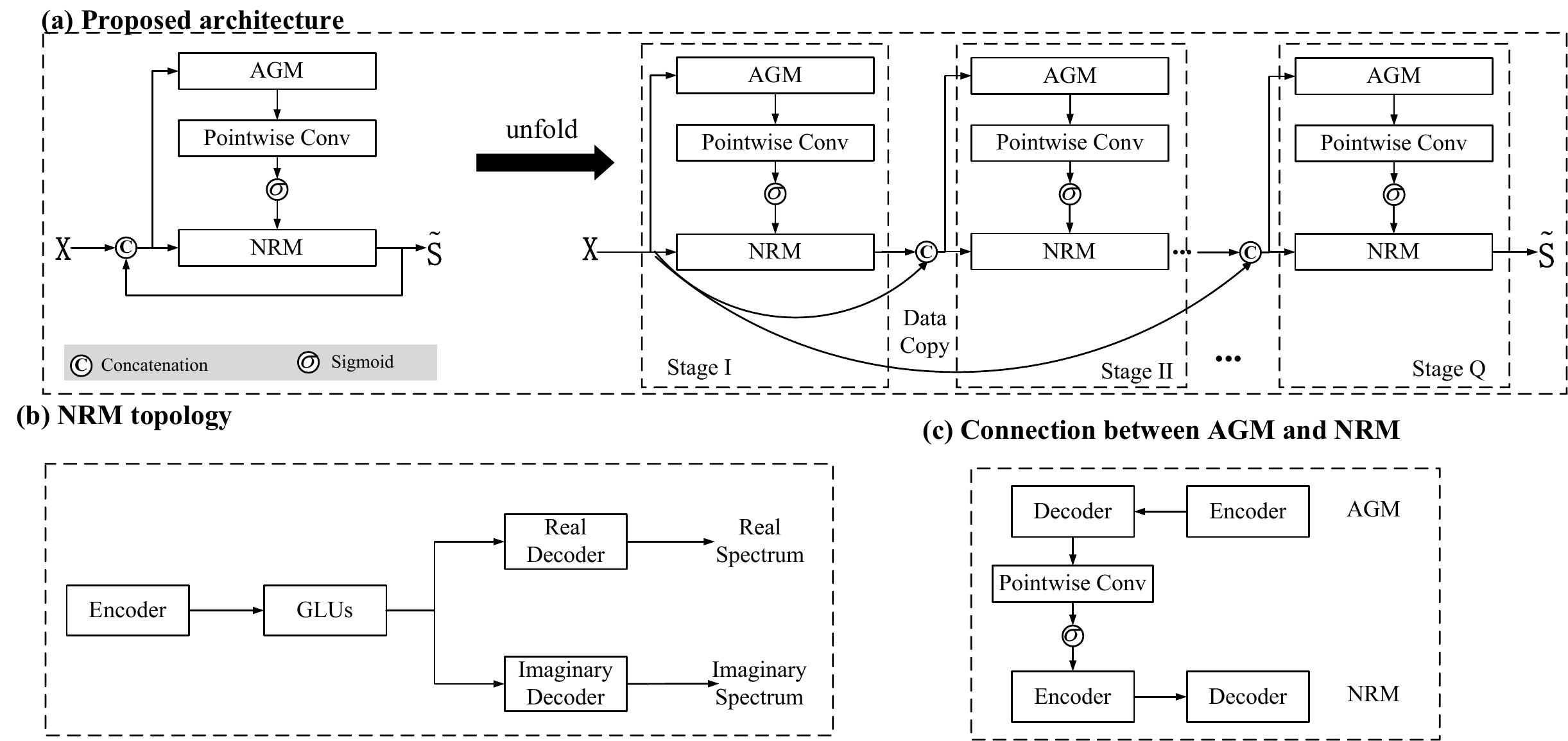}}
	\caption{The architecture of proposed model: (a) The model topology and its unfold format, (b) The topology of NRM, (c) The connection between NRM and AGM.}
	\label{fig:proposed-architecture}
\end{figure*}
\subsection{Recursive learning}
\label{recursive-learning}
The model used herein is based on a recently proposed time-frequency (T-F) network called DARCN~{\cite{li2020recursive}}, where recursive learning and dynamic attention are combined together with an organic method. Recursive learning is a type of learning protocol where the whole mapping procedure is decomposed into multiple stages and a memory mechanism is utilized to aggravate the information across different stages~{\cite{li2020time}}. To decrease the number of trainable parameters, the network parameters are shared among stages~{\cite{li2020speech}}. Fig.~{\ref{fig:proposed-architecture}}~(a) shows the architecture of recursive learning. Assuming the original noisy spectrum and the estimated spectrum in $l^{\rm th}$ stage are $\mathbf{X}$ and $\mathbf{S}^{l}$, respectively, the formulation in $l^{\rm th}$ stage is given as follows:
\begin{gather}
\label{eqn:equa1}
\mathbf{\tilde{S}}^{l} = \mathcal{F} \left(\mathbf{X}, \mathbf{\tilde{S}}^{l-1} \right),
\end{gather}
where $\mathcal{F}$ refers to the transform function of the network. Note that the training procedure is similar to deep unfolding, which can interpret the optimization algorithm as a recurrent neural network (RNN)~{\cite{wisdom2016deep, masuyama2019deep}}. We set the number of stages as 3 herein for the DNS Challenge.

\subsection{Dynamic attention}
\label{dynamic-attention}
In this submitted system, we design a type of unsupervised module called attention generator module (AGM) to dynamically adjust the feature distribution of the noise reduction module (NRM). This module simulates the auditory perception of the human being in complicated environments~{\cite{kaya2017modelling}}, i.e., when the received stimulus from the outside environment changes, the auditory attention tends to be modulated accordingly. In each stage, AGM receives the original noisy features and the estimation from the last stage and generates the attention distribution in the current stage. The distribution is then coupled with NRM via pointwise convolution and the Sigmoid function, which is shown in Fig.~{\ref{fig:proposed-architecture}}~(c). Assuming that the transform functions of AGM and NRM are $\mathcal{G}_{A}$ and $\mathcal{G}_{R}$, respectively, the formulation of the process is given as follows:
\begin{gather}
\label{eqn:equa2}
\mathbf{a}^{l} = \mathcal{G}_{A} \left( \mathbf{X}, \mathbf{\tilde{S}}^{l-1};\theta_{A} \right),\\
\mathbf{\tilde{S}}^{l} = \mathcal{G}_{R} \left( \mathbf{X}, \mathbf{\tilde{S}}^{l-1}, \mathbf{a}^{l} ;\theta_{R} \right),
\end{gather}
where $\theta_{A}$ and $\theta_{R}$ denote the parameters of AGM and NRM, respectively.

\subsection{Network parameter}
In this submitted system, we use a naive convolutional U-Net~{\cite{ronneberger2015u}} to model the attention generation process. It consists of 5 convolutional blocks in the encoder part and 5 deconvolutional blocks in the decoder part. Each of the blocks is comprised of a (de)convolutional layer, batch normalization, and exponential linear unit (ELU). The number of channels for each block is $\left(16, 32, 32, 32, 64, 64, 32, 32, 16, 16 \right)$. The kernel size is $\left( 2, 5 \right)$ along the temporal and frequency axis. As for noise reduction module, it encompasses three principal parts, namely stage recurrent neural network (SRNN)~{\cite{li2020time}}, attention U-Net (AU-Net)~{\cite{oktay2018attention}}, and multiple concatenated GLUs~{\cite{tan2018gated}}. SRNN is used to aggravate the information from previous stages and project these features into a higher feature space. The kernel setting in AU-Net is similar to AGM and the number of channels for the encoder and decoder are $\left( 16, 16, 32, 32, 64, 64 \right)$ and $\left( 64, 32, 32, 16, 16, 1 \right)$, respectively. GLUs are inserted in the middle of AU-Net, which consists of 6 GLU blocks to model the contextual correlations. The parameter detail of GLU is similar to~{\cite{tan2018gated}} except that the kernel size is set to 11. Note that different from directly mapping magnitude spectrum in~{\cite{li2020recursive}}, we take complex spectral mapping scheme, i.e., the real and the imaginary parts are estimated simultaneously. As Fig.~{\ref{fig:proposed-architecture}}~(b) shows, we design a separate decoder for real and imaginary components, respectively. To follow the rules of the DNS Challenge, all the convolutional operations are designed with the causal principle despite that several future frames may be beneficial to improve the performance of monaural speech enhancement~{\cite{cohen2005relaxed, hirszhorn2012transient}}. The total number of trainable parameters is 1.41 million, and the computational complexity to process one frame is 46.02 million floating-point fused multiply-adds, where the frame length is 20ms and the frame shift is 10ms.

\subsection{Dataset preparation and network training}
We use the provided scripts~{\footnote{https://github.com/microsoft/DNS-Challenge}} to generate the noisy-clean pairs for network training and validation. The provided clean dataset~{\footnote{https://github.com/microsoft/DNS-Challenge/tree/master/datasets}} is randomly split into two parts, where one contains 56200 clips and the other has 9148 clips. The two parts are used to generate the dataset for training and validation with the given noise dataset (including around 60000 clips). During the generation, each pair is fixed to 8 seconds. The range of signal-to-noise ratio (SNR) is set to range from $0\rm{dB}$ to $40\rm{dB}$ with the interval $2\rm{dB}$. The total length for training and that for validation are about 300 hours and 50 hours, respectively.

All the utterances are sampled at 16 kHz. 20ms Hamming window is applied with 50\% overlap between adjacent frames. To extract the T-F complex spectral features for both input and output, 320-point FFT is applied, leading to a 161-D feature vector. We concatenate the real and the imaginary parts along the channel axis. In the training stage, the initialized learning rate is set to 0.0002, we halve the learning rate if consecutive 3 validation loss increases, and we terminate the training if consecutive 5 validation loss increases. The minibatch is set to 4 at the utterance level.

\begin{table}[t]
	\caption{Objective results comparison between NSNet and DARCCN for synthesized data in $Test1$. PESQ and STOI are utilized as the metrics.}
	\centering
	\tiny
	\resizebox{0.45\textwidth}{!}{
		\begin{tabular}{cccc}
			\toprule
			&\multicolumn{1}{c}{Metrics} &PESQ  & STOI (in \%)\\
			\midrule
			\multirow{3}*{No Reverb}
			& Noisy &2.45 &91.52\\
			& NSNet &2.66 &77.62\\
			& DARCCN &3.17 &95.34\\
			\midrule
			\multirow{3}*{Reverb}
			&	Noisy &1.76 &56.01\\
			&	NSNet &1.77 &40.68\\
			&	DARCCN &1.88 &56.19\\
			\bottomrule
	\end{tabular}}
	\label{tbl:results1}
\end{table}

\section{Evaluation}
\label{experimental-evaluation}
The DNS Challenge provides two types of the test sets, both the first and the second test consist of 150 synthesized clips without reverberation, 150 synthesized clips with reverberation, and 300 real-recordings. Only the second test set is the blind test set used for the final comparison. We notate the two datasets as $Test1$ and $Test2$, respectively. The DNS Challenge also provides the baseline called NSNet~{\cite{9054254}}, and the model submitted to this challenge is named as DARCCN.

\subsection{Objective results comparison}
Since only synthesized clips in $Test1$ have the corresponding clean versions, we compare the results between the baseline and DARCCN with objective metrics, including PESQ scores~{\cite{recommendation2001perceptual}} and STOI scores~{\cite{taal2010short}}). The results are given in Table~{\ref{tbl:results1}}. One can get that the proposed model obtains consistently better performance than NSNet. For example, when going from NSNet to the proposed model in the reverberation free scenario, 0.51 PESQ improvement and 17.72\% STOI improvement are achieved. It indicates the superior performance of the proposed network. In noisy and reverberant cases, both models degrade their performance significantly. Nevertheless, the performance of our system is still acceptable whilst notable deterioration in STOI can be observed for NSNet. The performance degradation of the proposed system in reverberant cases can be explained clearly, this is because we do not consider the reverberation effect during the training process, and the network can not expect to effectively recognize and suppress the reverberation components.

\begin{table}[t]
	\caption{MOS scores between NSNet and the proposed system in $Test1$.}
	\centering
	\tiny
	\resizebox{0.45\textwidth}{!}{
		\begin{tabular}{cccccc}
			\toprule
			&Method &No Reverb  & Reverb & Real & Overall\\
			\midrule
			& Noisy &3.02 &2.44 &3.01 &2.87\\
			& NSNet &3.14 &2.16 &2.96 &2.81\\
			& DARCCN &3.64 &2.57 &3.22 &3.17\\
			\bottomrule
		\end{tabular}}
	\label{tbl:results2}
\end{table}

\begin{table}[t]
	\caption{MOS scores between NSNet and the proposed system in $Test2$.}
	\centering
	\tiny
	\resizebox{0.45\textwidth}{!}{
		\begin{tabular}{cccccc}
			\toprule
			&Method &No Reverb  & Reverb & Real & Overall\\
			\midrule
			& Noisy &3.32 &2.78 &2.97 &3.01\\
			& NSNet &3.49 &2.64 &3.00 &3.03\\
			& DARCCN &3.81 &3.02 &3.33 &3.37\\
			\bottomrule
	\end{tabular}}
	\label{tbl:results3}
\end{table}

\subsection{Subjective results comparison}
ITU-T P808 Subjective Evaluation of Speech Quality$\footnote{https://github.com/microsoft/P.808}$ is utilized to evaluate the Mean Opinion Score (MOS) of different speech enhancement (SE) models. Tables~{\ref{tbl:results2}} and~{\ref{tbl:results3}} present the subjective results for $Test1$ and $Test2$, which are provided by the organizers. Seen from the results, one can observe that compared with the baseline, our submitted system achieves consistently better subjective scores in both two test sets. For example, when going from NSNet to DARCCN, 0.36 and 0.34 average MOS improvements are obtained for $Test1$ and $Test2$, respectively, which verify the superiority of our submitted system in improving the subjective quality of enhanced utterances.

\begin{table}[t]
	\caption{Processing latency of proposed system per frame.}
	\centering
	\tiny
	\resizebox{0.35\textwidth}{!}{
		\begin{tabular}{cccc}
			\toprule
			&Method &GPU (ms)  & CPU (ms)\\
			\midrule
			& DARCCN &0.14 &5.43\\
			\bottomrule
		\end{tabular}}
	\label{tbl:results4}
	\vspace{-0.3cm}
\end{table}

\subsection{System latency}
Table~{\ref{tbl:results4}} summarizes the average processing latency of the proposed system for each frame. When measuring the latency time, 500 utterances are randomly chosen for processing. After the total processing time is obtained automatically, we calculate the average processing time for each frame. The experiments are repeated for 5 times. For the GPU configuration, we test the system on NVIDIA GeForce GTX 1660Ti with Max-Q Design. For the CPU configuration, we test the system on an Intel Core i7-9750H processor.

\section{Discussions and conclusions}
This technical report presents the submitted system for the DNS Challenge from Institute of Acoustics, Chinese Academy of Sciences and the evaluation results are also presented. From the final blind test results, one can see that the submitted system works well for both synthesized and real-recording clips. According to the guidelines of the DNS Challenge, it seems that our submitted system also belongs to the RT track. According to our experience, there are many ways to improve the performance of the submitted system for practical applications. First, the frame shift of the submitted system is only 10ms, so we can use the future 3 frames' information to further improve the performance of the DARCCN. Second, to improve cross-corpus generalization, we can use the smaller frame shift~{\cite{pandey2020cross}}, although the computational complexity may be increased. Third, one can generate reverberant (noisy) speech signals to train the model to increase its performance in both noisy and reverberant environments. Fourth, some preprocessing and postprocessing schemes can be introduced to further suppress some unnaturally residual noise components and reduce speech distortion~{\cite{hu2013cepstrum, 4358019}}. At last, instead of using the mean square error loss function, some newly developed loss functions can be introduced~{\cite{li2020supervised}}. 

\bibliographystyle{IEEEtran}

\bibliography{mybib}
\end{document}